\documentclass[reprint,aps,prl,showpacs,twocolumn]{revtex4-1}

\pdfoutput=1
\usepackage{mathptmx}
\usepackage{epsfig,amsopn}
\usepackage{graphicx}
\usepackage{amsmath,amssymb}
\usepackage{natbib,color}
\usepackage{braket}
\usepackage{float}
\usepackage{enumerate}
\usepackage{soul}

\allowdisplaybreaks[1]

\usepackage{adjustbox}
\usepackage[flushleft]{threeparttable}
\usepackage{soul}

\usepackage[framemethod=TikZ]{mdframed}
\newcounter{theo}
\renewcommand{\thetheo}{\arabic{theo}}

\def\bea{\begin{eqnarray}}
\def\eea{\end{eqnarray}}

\begin{document}

\title{Doubly stochastic continuous time random walk}

\author{{\normalsize{}Maxence Arutkin$^{1}$}
{\normalsize{}}}
\email{maxence@tauex.tau.ac.il}

\author{{\normalsize{}Shlomi Reuveni$^{1,2}$}
{\normalsize{}}}
\email{shlomire@tauex.tau.ac.il}

\affiliation{\noindent \textit{$^{1}$School of Chemistry, Center for the Physics \& Chemistry of Living Systems, Tel Aviv University,}}

\affiliation{\noindent \textit{$^{2}$Sackler Center for Computational Molecular \& Materials Science, Ratner Institute for Single Molecule Chemistry, Tel Aviv University, 6997801, Tel Aviv, Israel}}

\date{\today}

\begin{abstract}
Since its introduction, some sixty years ago, the Montroll-Weiss continuous time random walk has found numerous applications due its ease of use and ability to describe both regular and anomalous diffusion. Yet, despite its broad applicability and generality, the model cannot account for effects coming from random diffusivity fluctuations which have been observed in the motion of asset prices and molecules. To bridge this gap, we introduce a doubly stochastic version of the model in which waiting times between jumps are replaced with a fluctuating jump rate. We show that this newly added layer of randomness gives rise to a rich phenomenology while keeping the model fully tractable --- allowing us to explore general properties and illustrate them with examples. In particular, we show that the model presented herein provides an alternative pathway to Brownian yet non-Gaussian diffusion which has been observed and explained via diffusing diffusivity approaches.
\end{abstract}

\maketitle

In heterogeneous environments, e.g., porous media and biological cells, the random motion of molecules and particles may deviate from normal diffusion in different ways~\cite{havlin1987diffusion,bouchaud1990anomalous,metzler2000random,sokolov2012models,barkai2012single}. In particular, a series of recent experiments and computer simulations exhibit a similar pattern of anomalous behaviour: while the mean squared displacement is linear at all timescales, the displacement distribution is  Gaussian only for very long times or not at all (Fickian yet non-Gaussian)~\cite{wang2009anomalous,wang2012brownian,kim2013simulation,guan2014even,acharya2017fickian,chakraborty2019nanoparticle,chakraborty2020disorder}. 

An explanation to this unexpected behaviour, which seemingly breaks the central limit theorem, was given through the concept of superstatistics ~\cite{beck2001dynamical,beck2003superstatistics,wang2012brownian}. This idea describes a complex medium as one which has a distribution of potential diffusion coefficients. The displacement probability distribution of particles is then built by averaging over many diffusion pathways, each of which draws its particular diffusion coefficient from a distribution prescribed by  the complex medium. This approach can explain the observed experimental phenomenon of a system having a mean square displacement which is linear in time, alongside a non-Gaussian displacement probability distribution. Specifically, the latter is a weighted average (over the distribution of the diffusion coefficients) of Gaussian displacement probability distributions. Yet, the superstatistics approach cannot explain observed crossovers and transitions, e.g., from non-Gaussian distributions at short timescales to a Gaussian distribution at long timescales~\cite{metzler2020superstatistics}.

To address this issue, the concept of diffusing diffusivity was  introduced by Chubinsky and Slater~\cite{chubynsky2014diffusing} and further developed and explored by many others~\cite{jain2016diffusion,jain2017diffusing,tyagi2017non,chechkin2017brownian,sposini2018random,lanoiselee2018model,lanoiselee2018diffusion,thapa2018bayesian,wang2020unexpected,nampoothiri2021polymers,pacheco2021convergence,nampoothiri2022brownian,marcone2022brownian,alexandre2022non}. The basic idea is to describe the diffusion of particles with a diffusion equation in which the diffusion coefficient is itself diffusing. The Gaussian distribution at long timescales is then found universal for a diffusivity that is self-averaging in time. The displacement distribution at short timescales is, on the contrary, non-universal and depends on the equilibrium distribution of the diffusion coefficient. We note that similar phenomenology was observed  in the field of quantitative finance where the log-returns of stock prices diffuses with volatility that is analogous to the diffusion coefficient. As this volatility also diffuses in time~\cite{dragulescu2002probability,silva2004exponential,bouchaud2003theory}, the log-return exhibits similar transitions between short and long-time behaviour.

\begin{figure}[t!]
\includegraphics[width=0.45\textwidth]{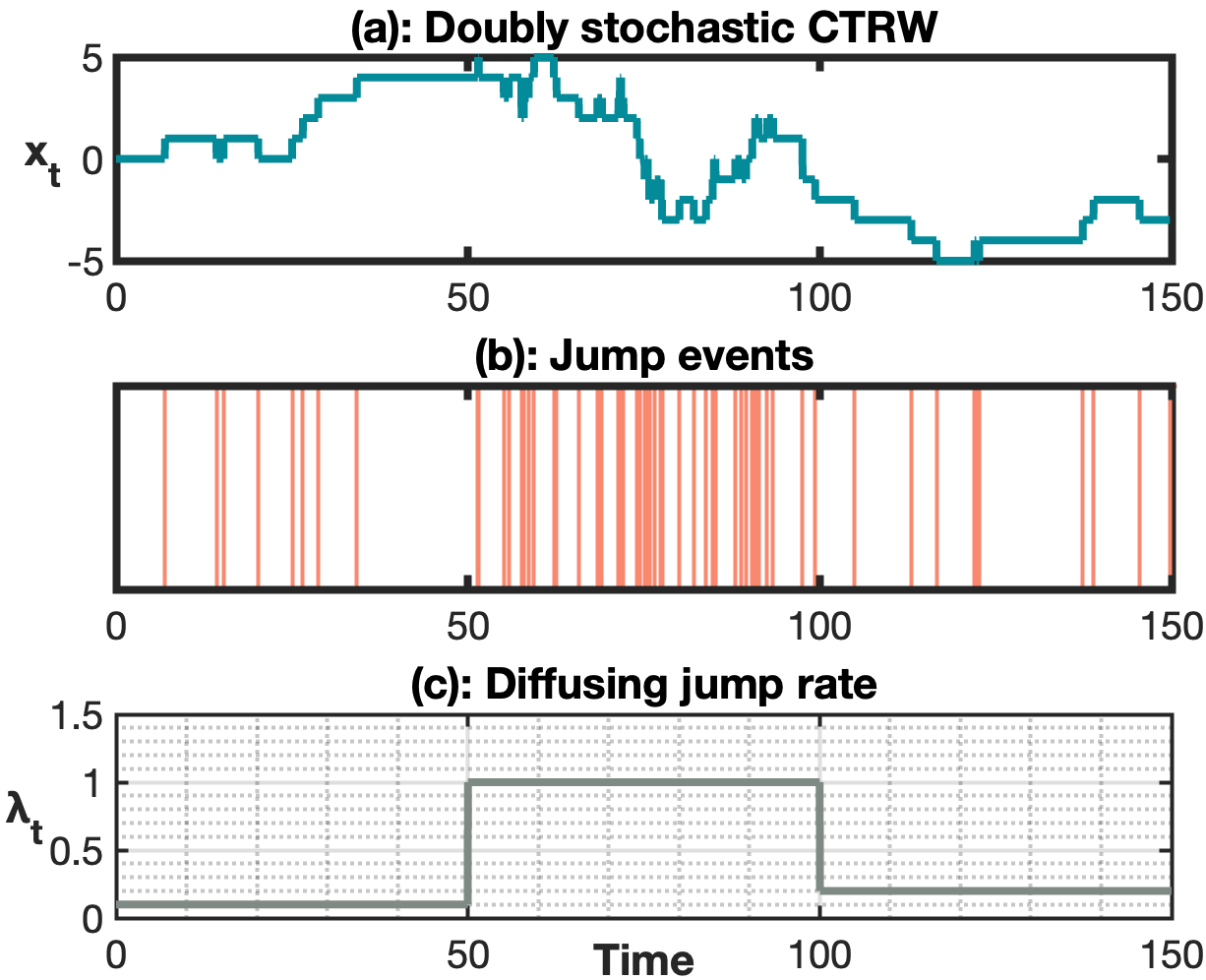}
\caption{The doubly stochastic CTRW has three layers of randomness. The fluctuating jump rate $\lambda_t$ in panel (c) gives the probability $\lambda_t dt$ for a jump to occur in the time interval $\left[t,t+dt\right]$. When a jump event occurs, bar-coded in panel (b), the random walker draws a random step and jumps to a new position as shown in panel (a). Here, we illustrate this general modeling framework using a jump rate $\lambda_t$ that is redrawn, every $\tau_r=50$ units of time, from an exponential distribution with unit mean. In this example, jumps are assumed symmetric, with equal probability to make a right/left step ($\pm1$). The resulting process exhibits periods of low and high activity, which give rise to deviations from Gaussian diffusion as observed in  heterogeneous environments. } 
\label{fig1}
\end{figure}

Diffusion can be seen as a continuum limit of a large class of random walk processes that occur on the microscopic scale. Yet, a general description of random walks with ``diffusing diffusivity" is currently missing from the physics literature. To bridge this gap, we introduce a new modelling framework that sets the foundations for the study of the diffusing diffusivity phenomenology from a random walks perspective. We start from the widely applied  continuous time random walk (CTRW)~\cite{montroll1965random,kutner2017continuous,shlesinger2017origins,klafter2011first}, where the walker jumps instantaneously from one position to another following a waiting period. In its simplest form, which gives rise to normal diffusion, waiting times in the CTRW are taken from an exponential distribution that is characterized by a constant jump rate $\lambda$. To introduce the equivalent of a diffusing diffusivity, we instead consider a diffusing jump rate $\lambda_t$~\cite{cox1955some,eliazar2005anomalous}, such that the probability to make a jump during the time interval $[t,t+dt]$ is given by $\lambda_t dt$. This results in a doubly stochastic version of the continuous time random walk. Next, we show that this newly added layer of randomness gives rise to a rich phenomenology which captures, inter alia, both regular and anomalous diffusion. The analysis brought below will be focused on transport and first-passage properties, leaving the correlation structure of the process to future study.

\emph{Doubly stochastic continuous time random walk (DSCTRW).---}The DSCTRW is schematised in Fig.~\ref{fig1}. A random walker takes steps at  random times which are determined by a Poisson process whose jump rate $\lambda_t$ is itself a random function of time. This doubly stochastic Poisson process is also known as a Cox process~\cite{cox1955some}. When a jump event occurs, the random walker takes a step $X$ from a distribution whose characteristic function is $\hat \phi\left(k\right)=\langle e^{ikX}\rangle $. We note that in actuarial science an equivalent model, the doubly stochastic compound Poisson process, has been studied \cite{bouzas2004theoretical,peters2011analytic,leveille2018conditional}. Also, the DSCTRW should not be confused with the noisy continuous time random walk \cite{jeon2013noisy} which is a different model.

The random walker's displacement is determined by $P(x,t)$, the probability to find it in position $x$ at time $t$. To compute this probability, let us first assume a given path for the jump rate process $\lambda_t$. The resulting jump process is then a time inhomogeneous Poisson process, and therefore the number of steps made until time $t$ will be given by the Poisson distribution with mean $\Lambda_t=\int_0^t\lambda_{t'}\text{d}t'$~\cite{kingman1992poisson,gallager2013stochastic}. Letting $\chi_n(t)$ denote the probability that the random walker made exactly $n$ steps until time $t$ we thus have $\chi_n(t)=\frac{\Lambda_t^n}{n!}e^{-\Lambda_t}$. The Fourier transform of the probability distribution of the displacement, \emph{conditioned on a given path of the diffusing jump rate}, reads
\begin{eqnarray}
\hat P\left(k,t\vert \lambda_t\right) = \sum_{n=0}^{+\infty}\chi_n(t)\hat\phi(k)^n = e^{-(1-\hat\phi(k))\Lambda_t}. 
 \label{Eq1:proba_cond}
\end{eqnarray}
Note, that this distribution only depends on the diffusing jump rate $\lambda_t$ through its time integral $\Lambda_t$. %Thus, the random walk dynamics does not depend on every detail of the diffusing jump rate but rather on its integrated properties.
Averaging Eq.~(\ref{Eq1:proba_cond}) with respect to the distribution of $\Lambda_t$, whose density we denote by $f(\Lambda_t)$, we obtain 
\begin{eqnarray}
\hat P(k,t) = \tilde \Lambda_t(1-\hat\phi(k)),
\label{Eq2:disp_fourier}
\end{eqnarray}
where $\tilde \Lambda_t(1-\hat\phi(k))=\int_0^\infty e^{-[1-\hat\phi(k)]\Lambda_t}f(\Lambda_t)\text{d}\Lambda_t$ is the Laplace transform of the integrated diffusing jump rate $\Lambda_t$ evaluated at $1-\hat\phi(k)$. Equation (\ref{Eq2:disp_fourier}) is the DSCTRW analog of the Montroll-Weiss formula. We emphasize its generality, and that it holds for both stochastic and deterministic paths of $\lambda_t$. 

\emph{Long-time behaviour.---} Starting from Eq.~(\ref{Eq2:disp_fourier}), we can understand both the short and long time behavior of the DSCTRW. Specifically, under mild assumptions, the long-time behavior is universally Gaussian. To show this, we need the following assumptions: (i) the mean and variance of the step distribution are finite; and (ii) the integrated jump rate has the following long-time asymptotics: $\Lambda_t=\int_0^t\lambda_{t'}\text{d}t' \simeq \overline\lambda t + \zeta_t$, where $\overline\lambda$ is the long time average of the fluctuating jump rate and where $\zeta_t$ is  Gaussian with $\langle \zeta_t \rangle = 0$ and $\langle \zeta^2_t \rangle = o(t^2)$. In particular, note that this condition holds for the typical case,  $\langle \zeta^2_t \rangle \sim t$, that arises (due to the central limit theorem) for a fluctuating jump rate that has a finite correlation time and a steady-state distribution with a finite mean and variance. Under these assumptions, the  Fourier transform of the scaled displacement converges to $\langle e^{ik \frac{x_t-\bar\lambda\langle X\rangle t }{\sigma(x_t)}} \rangle  \to e^{ -\frac{k^2}{2}}$~\cite{SupMat}, which is the Fourier transform of the standard normal. This result is analogous to the long time asymptotics that is obtained using the diffusing diffusivity approach.

The Gaussian limit above is universal given the aforementioned assumptions. Yet, we note that these can be broken in several different ways. For example, akin to scaled Brownian motion ~\cite{safdari2015aging,lim2002self,jeon2014scaled,he2008random}, one can think of cases where $\langle \lambda_t \rangle\propto \lambda t^{\alpha-1}$. These give $\langle \Lambda_t\rangle \propto \lambda t^{\alpha}$, which in turn yields sub-diffusion for $\alpha<1$ and super-diffusion for $\alpha>1$. However, unlike scaled Brownian motion, the DSCTRW also allows one to treat   processes where the mean and/or variance of the step size diverge. Diverging moments of the step distribution prevent moment expansion of $\hat\phi(k)$ and yield, in the  spirit of the generalized central limit theorem~\cite{levy1924theorie,klafter2011first}, $\alpha$-stable L\'evy asymptotics for the displacement. We illustrate this by taking steps from the Cauchy distribution, which returns a Cauchy distribution at long times (Fig. S1 in~\cite{SupMat}). A different way in which stable distributions may arise is when the jump rate has a steady-state with an infinite mean or variance. Such situations  give rise to stable distributions for $\Lambda_t$, and the position distribution can then be computed via  Eq.~(\ref{Eq2:disp_fourier}). The above-mentioned scenarios illustrate the richness of DSCTRW and highlight that this model can describe scenarios that fall outside the scope of the diffusing
diffusivity model.

\emph{Short-time behaviour.---}Contrary to long times, the short time displacement distribution is not universal and its shape depends on the steady-state distribution of the diffusing jump rate. Assuming the latter exists, we consider a case where the diffusing jump rate has been evolving for a very long time prior to the start of the experiment, such that it has converged to its steady-state.  In the short time limit $t\ll\tau_r $, with $\tau_r$ being the typical relaxation time of the diffusing jump rate process, the integrated rate behaves like $\Lambda_t=\int_0^t\lambda_{t'}\text{d}t' \simeq t \lambda_e$ with $\lambda_e$ drawn from the steady-state distribution whose Laplace transform we denote by $\tilde\lambda_e(s) = \langle e^{-s\lambda_e}\rangle$. 

Under these assumptions, the Laplace transform of the integrated rate can be expressed using the Laplace transform of the jump rate, $\tilde\Lambda_t(s) = \tilde\lambda_e(ts)$ and together with Eq.~(\ref{Eq2:disp_fourier}) we have
\begin{eqnarray}
\hat P(k,t) = \tilde\lambda_e\left(t(1-\hat\phi(k))\right).
\label{Eq3:disp_short_time}
\end{eqnarray}
Thus if we study the tail of this distribution in the case of a symmetric walk, we obtain $\hat P(k,t) \simeq \tilde\lambda_e(t k^2 \langle X^2 \rangle/2)$, which is a symmetric version of the rate probability distribution, that has no reason of being universal. For example, if $\lambda_e$ is exponentially distributed with mean $\overline\lambda$ -- which is the case for a jump rate process that is diffusing on $\mathbb{R}_+$ and with a negative drift pointing towards the origin --  we obtain $\hat P(k,t)\simeq\frac{1}{1+\overline\lambda t k^2 \langle X^2 \rangle/2}$ which is the Fourier transform of the Laplace distribution, also known as the bi-exponential. The probability distribution is then given by $ P(x,t) = \frac{1}{2\sigma_t}e^{-\frac{\vert x\vert}{\sigma_t}},$ with $\sigma_t = \sqrt{\overline\lambda t\langle X^2 \rangle/2}$. This  result is not universal, but noteworthy since the central limit theorem asserts that drift-diffusion often provides an excellent approximation to a host of biased random walk  processes that may govern the fluctuating jump rate. This could, perhaps, explain the prevalence of exponential tails observed experimentally at short times. For an alternative explanation see Refs. \cite{barkai2020packets, wang2020large}.

\emph{Moments.---}Moments in the DSCTRW can be computed by taking derivatives of
Eq.~(\ref{Eq2:disp_fourier}):  $\langle x_t^n\rangle = \lim_{k \to 0} (-i)^n\frac{\text{d}^n \hat P(k,t)}{\text{d}k^n}$. Specifically, we find that while the displacement probability distribution displays a non-Gaussian to Gaussian transition, the MSD of an unbiased walk is linear at all times $\langle x_t^2\rangle  = \langle X^2\rangle\langle \Lambda_t\rangle=\langle X^2\rangle\overline\lambda t$ ~\cite{SupMat}, provided $\lambda_t$ starts equilibrated and has a finite mean. More generally, the time dependence of the MSD is completely determined by the first moment of the integrated diffusing jump rate process and the second moment of the jump. The fourth moment of a symmetric DSCTRW is given by $\langle x_t^4\rangle=\langle X^4\rangle\langle \Lambda_t\rangle + 3\langle X^2\rangle^2 \langle \Lambda_t^2\rangle$~\cite{SupMat}. 

When bias is introduced to the random walk, the variance in position is given by $\sigma^2(x_t)=\langle X^2 \rangle \langle \Lambda_t \rangle + \langle X \rangle^2 (\langle \Lambda_t^2 \rangle - \langle \Lambda_t \rangle^2)$~\cite{SupMat}. Thus, when the integrated rate is stochastic, its variance does not vanish, resulting in additional dispersion. We illustrated this for a biased version of the toy model in Fig.~\ref{fig1}. There the mean and variance of the integrated jump rate are both linear in time, which leads to an apparent diffusion coefficient that can be significantly larger than in the absence of bias (Fig. S2 in SM~\cite{SupMat}). This enhanced dispersion, can be traced to the inherent coupling between drift and diffusion in the DSCTRW. Notably, the effect is absent from the diffusing diffusivity model, where drift and diffusion are completely decoupled.

\emph{An exactly solvable DSCTRW.---} To further illustrate the DSCTRW, we return to the model illustrated in Fig.~\ref{fig1}. There the diffusing jump rate is drawn every $\tau_r$ from an exponential distribution with mean jump rate $\bar\lambda=1$. The resulting jump rate is piece-wise constant, i.e., changes abruptly when a new time window starts. This can be seen as a simplified picture of more complicated diffusing jump rate processes whose autocorrelation time, $\tau_r$, is finite.

\begin{figure}[t!]
\includegraphics[width=0.45\textwidth]{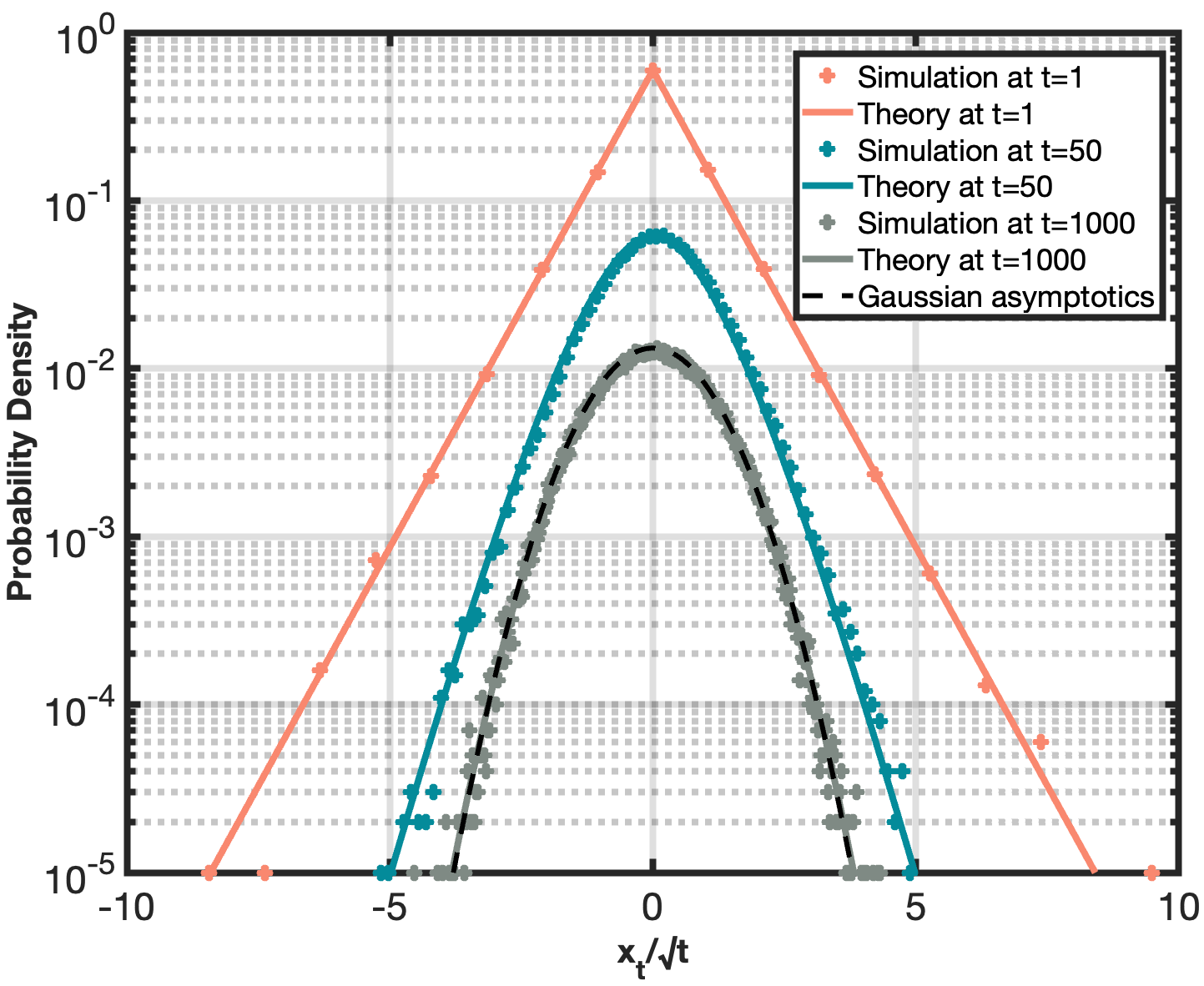}
\caption{Displacement probability distributions for the DSCTRW illustrated in Fig.~\ref{fig1}. Circles come from simulations and bold lines give analytical results obtained by inverting the Fourier transform in Eq.~(\ref{Eq6:disp_fourier_model}). Here, the mean jump rate was set to unity and its relaxation time to $\tau_r=10$. The distributions are plotted for three different times: $t=0.1\tau_r$ (orange) that exhibits a Laplace distribution, $t=5\tau_r$ (cyan) for which we see the beginning of a transition towards the Gaussian distribution, and $t=100\tau_r$ (gray) where the displacement distribution has converged to the Gaussian.}
\label{fig2}
\end{figure}

To get the propagator using Eq.~(\ref{Eq2:disp_fourier}), we compute the Laplace transform of the integrated jump rate, which in our model is a sum of independent random variables $\Lambda_t = \sum_{i=1}^{n_{\tau_r}}\lambda_{i}\tau_r +\lambda_{n_{\tau_r+1}}\delta_t$ with $n_{\tau_r}=\left \lfloor \frac{t}{\tau_r} \right \rfloor$ and $\delta_t = t-n_{\tau_r}\tau_r $. Therefore the Laplace transform, in  Eq.~(\ref{Eq2:disp_fourier}) is a  product of Laplace transforms of exponentially distributed random variables 
\begin{eqnarray}
 \hat{P}(k,t) = \tilde\lambda_e\left((1-\hat\phi(k))\tau_r\right)^{n_{\tau_r}}\tilde\lambda_e\left((1-\hat\phi(k))\delta_t \right),
 \label{Eq6:disp_fourier_model}
\end{eqnarray}
\noindent with $\tilde\lambda_e(s)=1/(1+ \bar\lambda s)$. Taking the walk to be simple symmetric, we have $\phi(k)=\cos(k)$, and a Laplace to Gaussian transition is observed when going from the short-time limit $t\ll\tau_r$ to the long-time limit $t\gg\tau_r$ (Fig.~\ref{fig2}).

%%%%%%%%%%%%%%%%%%%%%%%%%%%%%%%%%%%%%%%%%
%%%%%%%%%%%%%%%%%%%%%%%%%%%%%%%%%%%%%%%%%%

\emph{First passage statistics.---}The random time at which a stochastic process reaches a certain threshold, e.g. the encounter time of two molecules or the time a stock hits a certain price, can trigger a series of events. Thus, understanding the properties of first passage times is key to the explanation of many phenomena in statistical physics, chemistry, and finance~\cite{redner2001guide,metzler2014first,bray2013persistence}. To show how first-passage problems are solved in the context of the DSCTRW, we continue with the model illustrated in Fig.~\ref{fig1} and derive its first exit time  from an interval. This exit time exhibits interesting phenomenology due to the competition between different timescales. 

To obtain the exit time, we apply two simple steps: we first calculate the propagator conditioned on the number of steps taken, and then translate steps to time by summing over the corresponding probabilities $\chi_n(t)$. The propagator of the simple symmetric random walk inside an interval $[0,L]$ with absorbing boundaries is given by~\cite{feller1991introduction}
\begin{eqnarray}
    P(x,n\vert x_0) = \frac{2}{L}\sum_{p=0}^{L}\eta_p^n\sin(\nu_p x_0)\sin(\nu_p x),
    \label{Eq7:interval_prop}
\end{eqnarray}  
where $x_0$ is the initial position, $\nu_p=\frac{p\pi}{L}$ and $\eta_p=\cos{\nu_p}$. In the following we set $ u_p(x_0,x) = \sin(\nu_p x_0)\sin(\nu_p x)$. Translating steps to time, averaging over $\Lambda_t$, and summing over all lattice sites inside the interval, we obtain the survival probability
\begin{eqnarray}
    S(t\vert x_0) = \frac{2}{L}\sum_{x=1}^{L-1}\sum_{p=0}^{L}u_p(x_0,x)\tilde\Lambda_t(1-\eta_p),
    \label{Eq8:survival}
\end{eqnarray}
where $\tilde\Lambda_t$ which previously appeared in Eq.~(\ref{Eq2:disp_fourier}) should now be evaluated at $1-\eta_p$ instead of $1-\hat\phi(k)$. Taking the negative time derivative of $S(t\vert x_0)$, we obtain the probability density of the exit time
\begin{eqnarray}
    f(t\vert x_0) = \frac{2\bar\lambda}{L}\sum_{x=1}^{L-1}\sum_{p=0}^{L}\frac{u_p(x_0,x)(1-\eta_p)\tilde\Lambda_t(1-\eta_p)}{1+ (1-\eta_p)\delta_t\bar\lambda }.
    \label{Eq10:fpt}
\end{eqnarray}

Analyzing the density in Eq.~(\ref{Eq10:fpt}) right before and just after integer multiples of $\tau_r$, we analytically show that it experiences jumps whose magnitude increases with  $\tau_r$~\cite{SupMat}. This feature is a manifestation of the sharp transitions experienced by the diffusing jump rate in our model (Fig.~\ref{fig1}c).  Indeed, particles that initially drew a smaller than average jump rate have a higher probability to survive simply by virtue of moving less. As a result, the system becomes enriched with such particles,  until a new jump rate is redrawn at $t=\tau_r$. This leads to a jumps in the first-exit density, which come from slowly moving particles that become faster moving and perish. 

\begin{figure}[t!]
\includegraphics[width=0.475\textwidth]{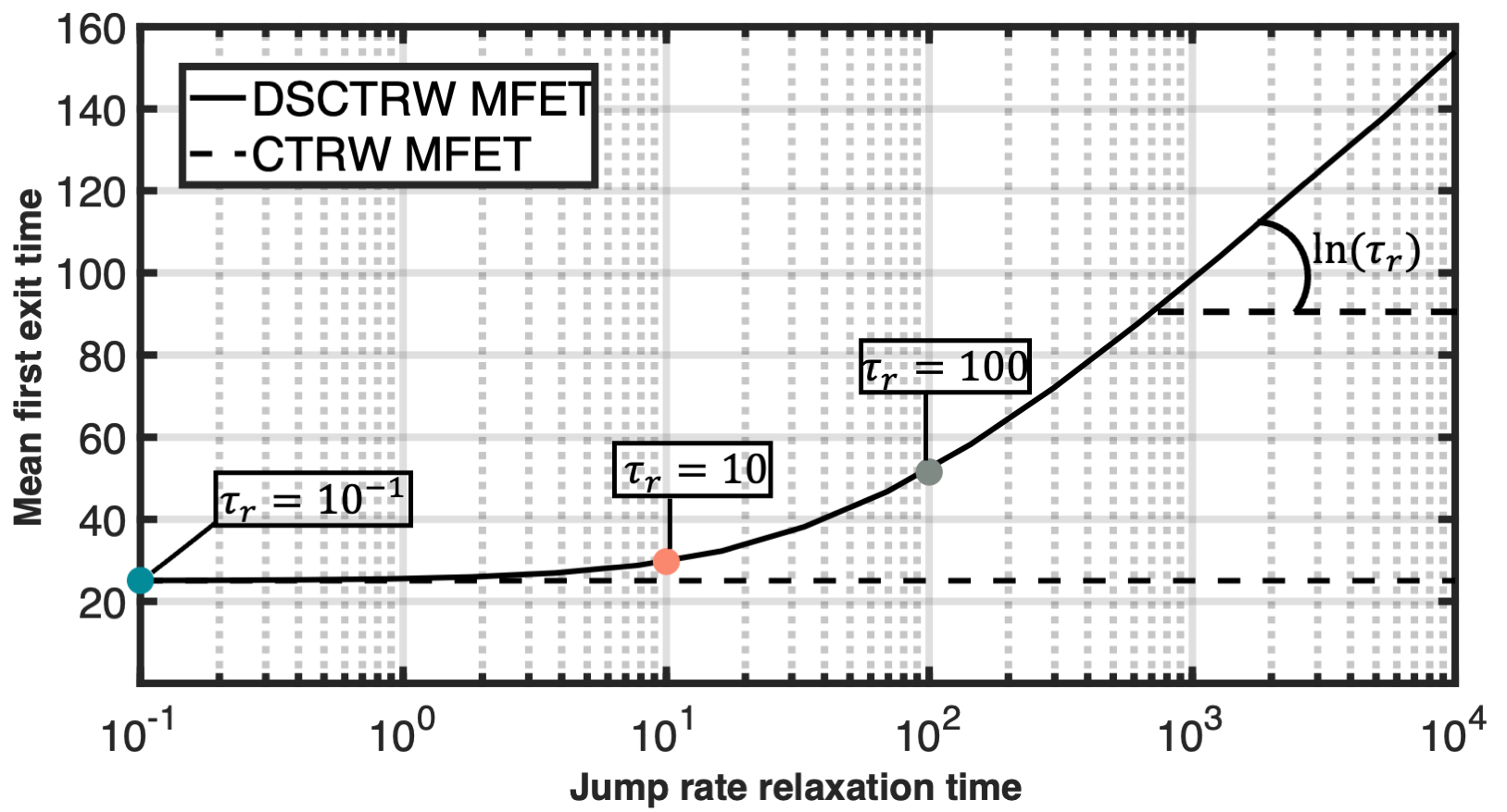}
\caption{Mean first exit time (MFET) from the interval $\left[0,10\right]$, starting  at $x_0=5$, for the DSCTRW illustrated in Fig.~\ref{fig1}. For fast jump rate relaxation (small $\tau_r$), we recover the MFET of a simple CTRW (dashed line) with a fixed jump rate corresponding to the mean jump rate, $\overline\lambda=1$, of the DSCTRW.  In the other extreme, i.e., for slow relaxation times (large $\tau_r$), the MFET scales as $\sim\ln(\tau_r)$. The bold line is plotted by averaging over Eq.~(\ref{Eq10:fpt}) and circles denote MFETs that correspond to the distributions in Fig. S3.}
\label{fig4}
\end{figure}

The jump rate relaxation time, $\tau_r$, also has a profound impact on the mean first exit time (MFET) from the interval, which we obtain by averaging over the density in Eq.~(\ref{Eq10:fpt}). Results are given in Fig.~\ref{fig4}, where we distinguish between two limiting behaviours. When relaxation times are fast, jump rate fluctuations play a lesser role as they are averaged over. In this limit, the jump rate can be seen as if it was fixed and equal to the mean jump rate $\overline\lambda$. The MFET can then be approximated as $\simeq(L-x_0)x_0/\overline\lambda$, which is the MFET of a simple CTRW with a fixed jump rate $\overline\lambda$. Situation is different for slow relaxation times $\tau_r\gg(L-x_0)x_0/\overline\lambda$. In this case, a typical particle is ``stuck'' with its initially drawn jump rate until it leaves the interval. The MFET can then be approximated by averaging $(L-x_0)x_0/\lambda$ over the initial distribution of the jump rate $\lambda$ (here assumed equivalent to the steady-state distribution). Doing so for an exponential distribution of jump rates leads to a logarithmic divergence, which we regularize by noting that the relaxation rate, $\tau_r^{-1}$, serves a lower cutoff for the integration. Indeed, very slow particles that do not leave the interval by the relaxation time, will consequently draw a new jump rate. More generally we find that the asymptotic dependence of the MFET on $\tau_r$ depends on the small rate asymptotics of steady-state distribution of the jump rate~\cite{SupMat}.   

%%%%%%%%%%%%%%%%%%%%%%%%%%%%%%%%%%

%%%%%%%%%%%%%%%%%%%%%%%%%%%%%%%%%%%%%%%%

We note that the first-passage results obtained above can be generalized for many other geometries and boundary conditions. This can be done whenever the propagator in real, or Fourier space, admits a standard  eigenmode expansion of the form which appears in Eq.~(\ref{Eq7:interval_prop}). Crucially, when terms in the series are proportional to $\eta_p^n$ , one can easily translate steps to time by summing over the corresponding probabilities $\chi_n(t)$ and taking expectations. For additional  examples that can be solved in a similar fashion see~\cite{giuggioli2020exact}, which gives exact results for propagators of random walks in confined geometries.

\emph{Conclusions.---}In this letter, we introduced a doubly stochastic version of the renowned CTRW. The model allows for a general description of a random walk which is driven by a time dependent jump rate that fluctuates randomly. Despite this added layer of complexity, the model remains fully tractable and we obtained a general formula for the displacement probability distribution. A rich phenomenology emerged from the analysis, asserting that the doubly stochastic continuous time random walk can be used to describe not only super, sub, and normal diffusion, but also  the Brownian yet non-Gaussian diffusion that has recently been observed in various systems. The tractability of the model further lends itself to the computation of first-passage times which---similar to the displacement distribution---display striking transitions as a function of the jump rate relaxation time. The random walk approach developed herein complements the diffusing diffusivity approach that was developed for Brownian motion, and further extends it by allowing for unlimited freedom in the interplay between the distribution of jumps and the properties of the fluctuating jump rate.  

\textit{Acknowledgments.}---M.A. acknowledges discussions with Samuel Gelman, Ulysse Mizrahi and Bara Levit. S.R. acknowledges support from the Israel Science Foundation (grant No. 394/19). This project  received funding from the European Research Council (ERC) under the European Union’s Horizon 2020 research and innovation program (Grant agreement No. 947731).

\bibliography{biblio.bib}

\end{document}

%%%%%%%%%%%%%%%%%%%%%%%%%%%%%%%%%%%%%%%%%%%%%%%%%%%%%%%%%%%%%%%%%%%%%%%%%%%%%%%%%%%%%%%%%%%%%%%%%%%%%%%%%%%%%%%%%%%%%%%%%
%\widetext

%\newpage
%\pagestyle{empty}

%%%%%%%%%% Merge with supplemental materials %%%%%%%%%%
%\pagebreak
%\begin{widetext}

%%%%%%%%%% Merge with supplemental materials %%%%%%%%%%
%%%%%%%%%% Prefix a "S" to all equations, figures, tables and reset the counter %%%%%%%%%%
\setcounter{equation}{0}
\setcounter{figure}{0}
\setcounter{table}{0}
\setcounter{page}{1}
\setcounter{section}{0}
\makeatletter
\renewcommand{\theequation}{S\arabic{equation}}
\renewcommand{\thefigure}{S\arabic{figure}}
\renewcommand{\thesection}{S\Roman{section}} 
\renewcommand{\bibnumfmt}[1]{[S#1]}
\renewcommand{\citenumfont}[1]{S#1}

\begin{center}\Large{Supplemental Material for: ``Doubly stochastic continuous time random walk''}\end{center}

\author{{\normalsize{}Maxence Arutkin$^{1}$}
{\normalsize{}}}
\email{maxence@tauex.tau.ac.il}

\author{{\normalsize{}Shlomi Reuveni$^{1,2}$}
{\normalsize{}}}
\email{shlomire@tauex.tau.ac.il}

\affiliation{\noindent \textit{$^{1}$School of Chemistry, Center for the Physics \& Chemistry of Living Systems, Tel Aviv University,}}

\affiliation{\noindent \textit{$^{2}$Sackler Center for Computational Molecular \& Materials Science, Ratner Institute for Single Molecule Chemistry, Tel Aviv University, 6997801, Tel Aviv, Israel}}

\maketitle
\tableofcontents

%%%%%%%%%%%%%%%%%%%%%%%%%%%%%%%%%%%%%%%%%%%%%%%%%%%%%%%%%%%%%%%%%%%%%%%%%%%%%%%%%%%%%%%%%%%%%
\section{Convergence towards the Gaussian at long times}
\noindent To prove convergence towards the Gaussian at long times, as discussed in the main text, we assume an unbiased random walk for which: (i) the variance of the step distribution is finite; and (ii) the integrated jump rate has the following long-time asymptotics: $\Lambda_t=\int_0^t\lambda_{t'}\text{d}t' \simeq \overline\lambda t + \zeta_t$, where $\overline\lambda$ is the long time average of the fluctuating jump rate and where $\zeta_t$ is Gaussian with $\langle \zeta_t \rangle = 0$ and $\langle \zeta^2_t \rangle = o(t^2)$. In particular, note that the proof provided below holds for the typical case,  $\langle \zeta^2_t \rangle \sim t$, which arises (due to the central limit theorem) for a fluctuating jump rate that has a finite correlation time and a steady-state distribution with a finite mean and variance.

\noindent To show convergence towards a Gaussian at long timescales, we consider the re-scaled process $x_t/\sqrt{ \langle X^2\rangle \bar\lambda t}$, whose characteristic function is given by

\begin{eqnarray}
    \langle e^{ik x_t/\sqrt{ \langle X^2\rangle \bar\lambda t}} \rangle = \tilde \Lambda_t\left(1-\hat \phi\left(\frac{k}{\sqrt{\langle X^2\rangle \bar\lambda t}}\right)\right),
\end{eqnarray}

\noindent by use of  Eq. (2) in the main text. We now utilize the fact that in the long time limit the integrated jump rate can be expressed as $\Lambda_t=\bar \lambda t +\zeta_t$, and its  Laplace transform is thus given by $\tilde\Lambda_t(s)=e^{-\bar\lambda t s}e^{+s^2\langle \zeta^2_t \rangle /2}$. To proceed, we note that for fixed $k$ and large $t$ one can approximate  $\hat\phi(k/\sqrt{\langle X^2\rangle \bar \lambda t})\simeq 1-\frac{k^2}{2 \bar \lambda t}$, which in turn gives 
\begin{eqnarray}
    \langle e^{ik x_t/\sqrt{ \langle X^2\rangle \bar \lambda t}} \rangle &\simeq & e^{-k^2/2 }\tilde \zeta_t(k^2/2 \bar \lambda t) \simeq  e^{- k^2/2 }\left(1+\frac{k^4}{8{\bar \lambda}^2 t^2}o(t^2)\right)\\
                & \rightarrow & e^{- k^2/2}~,
\end{eqnarray}

\noindent where we have used the assumption $\langle \zeta^2_t \rangle = o(t^2)$. As the characteristic function of the re-scaled process tends to the characteristic function of a Gaussian one can conclude that the long time asymptotic of the process is also Gaussian. 

The case of a biased random walk is slightly different. We examine the characteristic function of the displacement in the moving frame of the random walker's average position 
\begin{eqnarray}
    \langle e^{ik \frac{x_t-\bar\lambda\langle X\rangle t }{\sigma(x_t)}} \rangle = e^{-ik \bar\lambda\frac{\langle X\rangle t }{\sigma(x_t)}} \tilde \Lambda_t\left(1-\hat \phi\left(\frac{k}{\sigma(x_t)}\right)\right),
\end{eqnarray}

\noindent where $\sigma^2(x_t) =\langle X^2\rangle\langle \Lambda_t\rangle+\langle X\rangle^2\langle \Lambda_t^2\rangle-\langle X\rangle^2\langle \Lambda_t\rangle^2 \simeq\langle X^2\rangle\overline\lambda t+\langle X\rangle^2 \langle \zeta^2_t \rangle$ (see moments section below). To proceed, we note that for fixed $k$ and large $t$ one can approximate  $\hat\phi(k/\sigma(x_t))\simeq 1+ik\frac{\langle X\rangle}{\sigma(x_t)}-\frac{k^2\langle X^2\rangle}{2\sigma^2(x_t)}$, which in turn gives
\begin{eqnarray}
    \langle e^{ik \frac{x_t-\bar\lambda\langle X\rangle t }{\sigma(x_t)}} \rangle &\simeq & e^{-ik \bar\lambda\frac{\langle X\rangle t }{\sigma(x_t)}} e^{+i\bar\lambda t \frac{k\langle X\rangle}{\sigma(x_t)}-\bar\lambda t\frac{k^2\langle X^2\rangle}{2\sigma^2(x_t)} }e^{-\frac{k^2\langle X\rangle^2}{2\sigma^2(x_t)}\langle \zeta^2_t \rangle-i\frac{k^3\langle X\rangle\langle X^2\rangle}{2\sigma^3(x_t)}\langle \zeta^2_t \rangle+\frac{k^4\langle X^2\rangle^2}{8\sigma^4(x_t)}\langle \zeta^2_t \rangle}\\
                & \rightarrow &  e^{-\frac{k^2\langle X^2\rangle \bar\lambda t}{2\sigma^2(x_t)} }e^{-\frac{k^2\langle X\rangle^2 \langle \zeta^2_t \rangle}{2\sigma^2(x_t)}}= e^{-k^2/2~}.
\end{eqnarray}
\noindent In the above we have utilized the assumption that $\zeta_t$ is asymptotically Gaussian, and moreover noted that $\lim_{t\to\infty} \langle \zeta_t^2\rangle/\sigma^3(x_t)=\lim_{t\to\infty} \langle \zeta_t^2\rangle/\sigma^4(x_t)=0$.

%%%%%%%%%%%%%%%%%%%%%%%%%%%%%%%%%%%%%%%%%%%%%%%%%%%%%%%%%%%%%%%%%%%%%%%%%%%%%%%%%%%%%%%%%%%%%

\section{Breakdown of the long-time Gaussian asymptotics for the Cauchy random walk}

\begin{figure}[H]
        \centering
        \includegraphics[width=12cm]{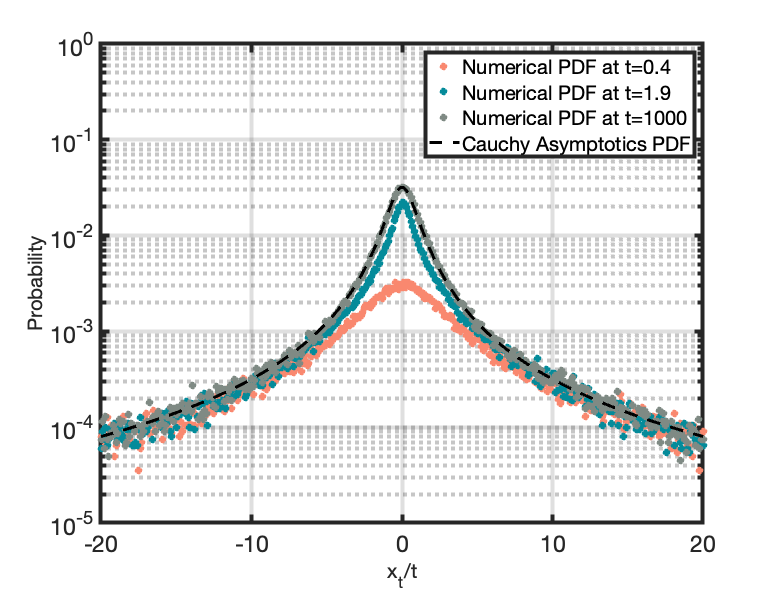}
                \caption{To demonstrate the breakdown of the long-time Gaussian asymptotics in situations where the mean and/or variance of the step size diverge, we consider the Cauchy random walk. To this end we take a random walker that draw its steps from a Cauchy distribution whose  characteristic function is $e^{-\vert k\vert}$. The fluctuating jump rate follows a similar dynamic as in the main text. Namely, every $\tau_r = 10$, a new jump rate is drawn from an exponential distribution with unit mean. Here, we plot the displacement probability distribution for  random walks that have made at least one step (note rescaling by $t$). The atoms at zero displacement, corresponding to random walks that did not move, are not shown.  The figure displays the evolution of the PDFs (scatter plots) for three time points: t = 0.4 (orange), t = 1.9 (cyan), and t = 1000 (gray). The black dashed line represents the the asymptotic Cauchy distribution $1/\pi(1+x^2)$, which is obtained by small $k$ expansion of Eq. (4) in the main text. The plots demonstrate the convergence of the PDFs towards the asymptotic PDF as time progresses.}
        \label{fig:Cauchy}
\end{figure}

\section{Moments of the doubly stochastic CTRW}
\noindent Moments in the DSCTRW can be computed by taking derivatives of
Eq. (2) in the main text
\begin{eqnarray}
\langle x_t^n\rangle = \lim_{k \to 0} (-i)^n\frac{\text{d}^n \hat P(k,t)}{\text{d}k^n}= \lim_{k \to 0} (-i)^n\frac{\text{d}^n \tilde \Lambda_t(1-\hat\phi(k))}{\text{d}k^n}. 
\end{eqnarray}
\noindent For example, the first moment is given by 
\begin{eqnarray}
    \langle x_t\rangle &=& \lim_{k\rightarrow 0} i\hat\phi'(k)\tilde\Lambda'_t(1-\hat\phi(k))\\
    &=&i\hat\phi'(0)\tilde\Lambda'_t(0)\\
    &=&\langle X\rangle\langle\Lambda_t\rangle
\end{eqnarray}
Similarly, the second moment is given by
\begin{eqnarray}
    \langle x_t^2\rangle = -\frac{\text{d}^2\hat P(k,t)}{\text{d}k^2}\vert_{k=0} = -\hat\phi''(0)\tilde\Lambda'_t(0) - \hat\phi'(0)^2\tilde\Lambda''_t(0) =\langle X^2\rangle \langle \Lambda_t\rangle + \langle X\rangle^2 \langle \Lambda_t^2\rangle,
\end{eqnarray}
which for an unbiased random walk boils down to  
\begin{eqnarray}
    \langle x_t^2\rangle = \langle X^2\rangle\langle \Lambda_t\rangle.
\end{eqnarray}
Thus, the MSD of an unbiased random walk depends on the second moment of the walk and the first moment of the integrated jump rate. Specifically, if $\lambda_t$ has a finite mean and it starts equilibrated at $t=0$, we have $\langle \Lambda_t \rangle=\langle \int_{0}^{t}\lambda_{t'}\text{d}t' \rangle =  \int_{0}^{t}\langle\lambda_{t'}\rangle\text{d}t'=  \overline\lambda t $ which yields a linear MSD at all times. Also, even if $\lambda_t$ is not equilibrated at time zero but it is self averaging around a mean value, a linear MSD will emerge following a relaxation period. Finally, we note that the fourth moment of an unbiased random walk reads
\begin{eqnarray}
    \langle x_t^4\rangle = \frac{\text{d}^4\hat P(k,t)}{\text{d}k^4}\vert_{k=0} = -\hat\phi''''(0)\tilde\Lambda'_t(0) + 3 \hat\phi''(0)^2\tilde\Lambda''_t(0) =\langle X^4\rangle\langle \Lambda_t\rangle + 3\langle X^2\rangle^2 \langle \Lambda_t^2\rangle,
\end{eqnarray}
\noindent where we have further assumed that the walk is not only unbiased but also unskewed: $\langle X^3\rangle=i\hat\phi'''(0)=0$. Clearly, the fourth moment depends on the second and fourth moment of the step distribution and the first two moments of the integrated jump rate. Deviations from Gaussianity can be quantified by looking at the kurtosis  $\kappa = \langle x_t^4\rangle/\langle x_t^2\rangle^2$,  which in the DSCTRW transitions from the kurtosis of the short time displacement distribution to the kurtosis of the normal distribution (equal to three).

%%%%%%%%%%%%%%%%%%%%%%%%%%%%%%%%%%%%%%%%%%%%%%%%%%%%%%%%%%%%%%%%%%%%%%%%%%%%%%%%%%%%%%%%%%%%%
\section{Enhanced dispersion of the biased DSCTRW}

\begin{figure}[!h]
        \centering
        \includegraphics[width=11cm]{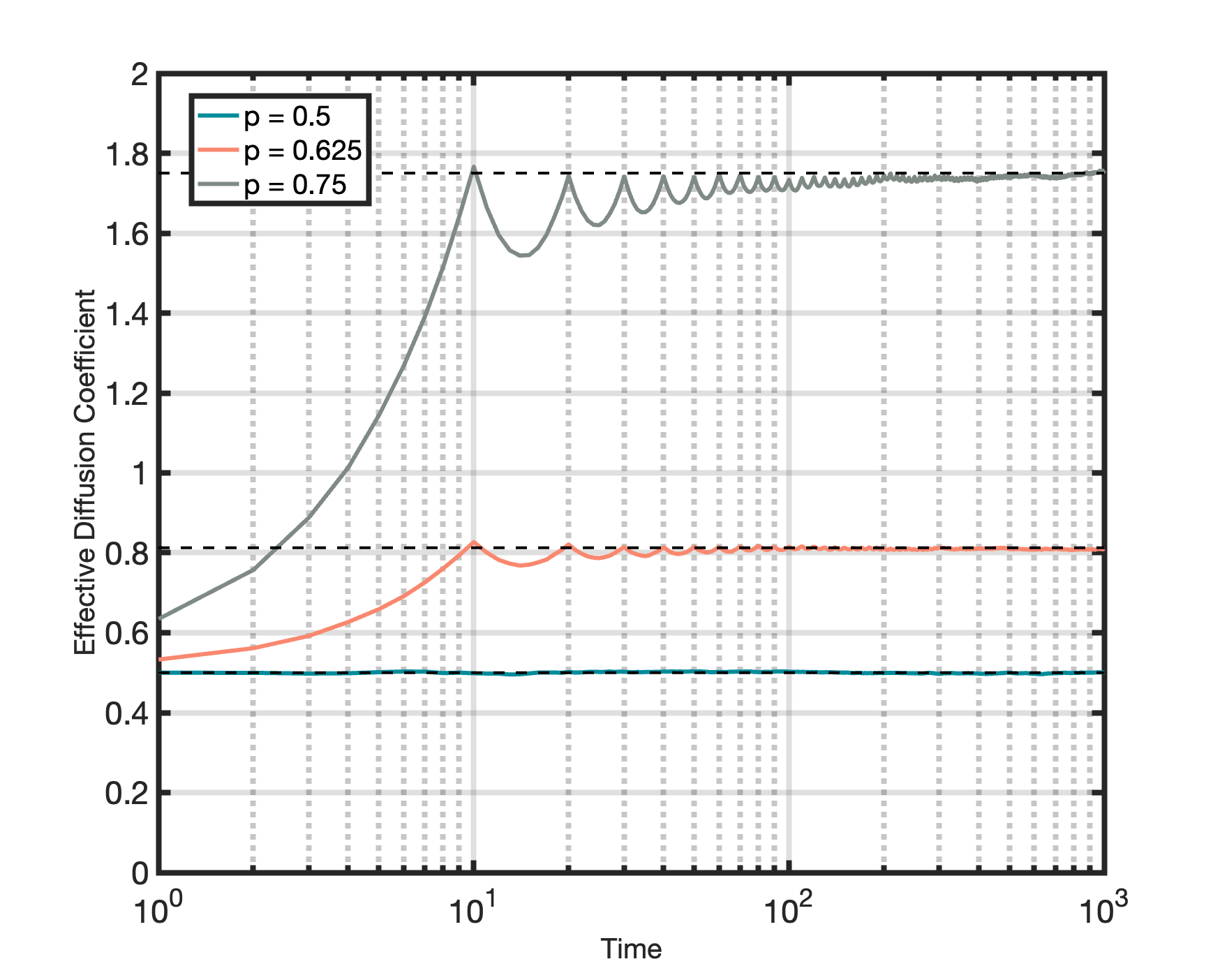}
        \caption{The effective diffusion coefficient from Eq. (\ref{eff_DC}) vs. time (logarithmic scale). Plots are made for three different DSCTRWs, each with a different probability $p$ to make a step to the right. The black dashed lines corresponds to asymptotic  effective diffusion coefficient in Eq.~\ref{Eq:SM_Deff}, with $\tau_r=10$. The results show that biased DSCTRWs are characterized by enhanced dispersion, i.e., have larger effective diffusion coefficients due to the inherent coupling between bias and diffusion in this model.}
        \label{fig:drift_dsctrw}
\end{figure}

\noindent To illustrate the enhanced dispersion of the biased DSCTRW, we combine Eqs. (S10) and (S11) to obtain the variance which reads
\begin{eqnarray}
    \sigma^2(x_t) = \langle X^2\rangle \langle \Lambda_t\rangle + \langle X\rangle^2 \left(\langle \Lambda_t^2\rangle- 
\langle \Lambda_t\rangle^2\right).
\label{Variance_SM}
\end{eqnarray}
From here, it is clear that for deterministic jump rates, the variance of the integrated jump rate vanishes and so is the second term in Eq. (\ref{Variance_SM}). In this scenario, the effective diffusion coefficient, 
\begin{equation}
D_{\text{eff}}(t)=\sigma^2(x_t)/2t, 
\label{eff_DC}
\end{equation}
does not depend on $\langle X\rangle$ and is the same for biased and unbiased random walks. However, for stochastic jump rates the variance of the integrated jump rate is always positive, which means that the effective diffusion coefficient can be significantly larger in the presence of bias. To exemplify this concretely, we consider once more the toy model in Fig. 1. We recall that in this model the integrated jump rate is given by $\Lambda_t = \sum_{i=1}^{n_{\tau_r}}\lambda_{i}\tau_r +\lambda_{n_{\tau_r+1}}\delta_t$, where $\{\lambda_1,\lambda_1,...,\lambda_{n_{\tau_r+1}}\}$ are independent identically distributed exponential random variables with mean jump rate $\bar\lambda=1$,  $n_{\tau_r}=\left \lfloor \frac{t}{\tau_r} \right \rfloor$, and $\delta_t = t-n_{\tau_r}\tau_r $. We thus have 

\begin{equation}
    \langle \Lambda_t \rangle =  t,
\end{equation}
and
\begin{equation}
    \sigma^2(\Lambda_t) = n_{\tau_r}\tau_r^2 + \delta_t^2.
\end{equation}

%\begin{equation}
%    \langle \Lambda_t^2 \rangle \sim 
%\lambda_0^2 t^2 + \lambda_0^2 t\tau_r,
% \end{equation}
\noindent With the  above, we a consider a random walk that has probability $p$ to make a step to the right and $1-p$ to make a step to the left. Combining Eqs. (S14-S16), we find that the  effective diffusion coefficient of this random walker is asymptotically given by 
\begin{eqnarray}    \lim_{t\to\infty}D_{\text{eff}}(t)=\lim_{t\to\infty}\frac{\sigma^2(x_t)}{2t}= \frac{1 + (2p-1)^2\tau_r}{2},
    \label{Eq:SM_Deff}
\end{eqnarray}
which for $p \neq 1/2$ is larger than half --- the effective diffusion coefficient of the unbiased walk. This result is further illustrated in Fig. S2.

\section{Discontinuities of the first passage density in Eq. (7)}

%%%%%%%%%%%%%%%%%%%%%%%%%%%%
\begin{figure}[h!]
\includegraphics[width=11cm]{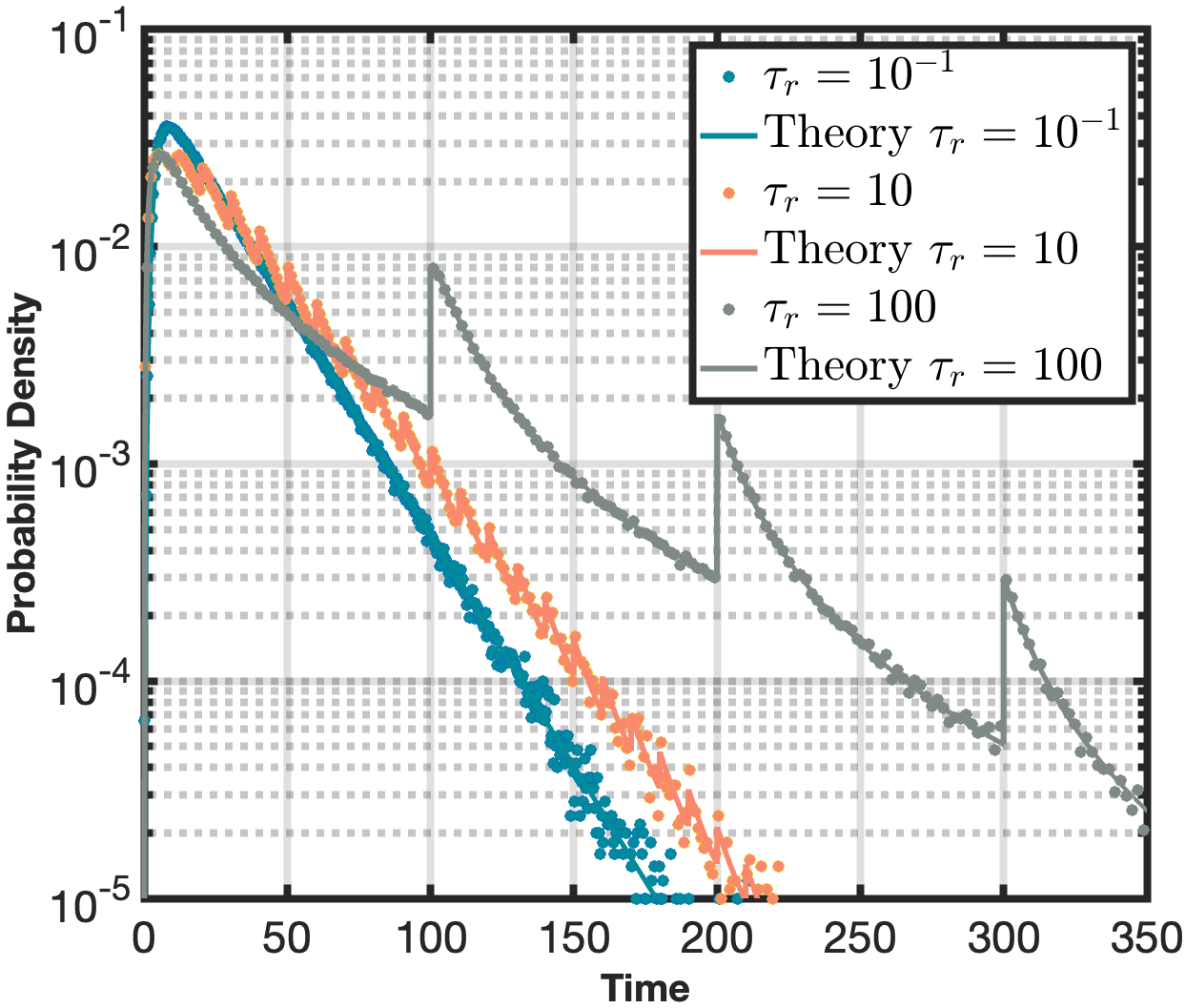}
\caption{First exit time distributions from the interval $\left[0,10\right]$, starting at $x_0=5$, for the DSCTRW illustrated in Fig. 1 of the main text. Results are plotted for three different jump rate relaxation times: $\tau_r = 10^{-1}$ (cyan), $\tau_r = 10^{1}$ (orange), and $\tau_r = 10^{2}$ (gray). Bold lines are plotted using Eq. (7) of the main text, and circles come from numerical simulations.}
\label{fig3}
\end{figure}
%%%%%%%%%%%%%%%%%%%%%%%%%%%%

\noindent In Fig. S3, we observe discontinuities -- jumps -- in the first passage density localized at integer multiple of $\tau_r$. This is a manifestation of the sharp transition experienced by the diffusing jump rate of the DSCTRW illustrated in Fig. 1. To analytically prove the existence of discontinuities we need to show that at times which are integer multiples of $\tau_r$ the first-passage density has different limits when approached from the left/right. We will show this for $t=\tau_r$, but the proof is identical for any integer multiple of $\tau_r$. We set $\lim_{\epsilon\to 0} f(\tau_r - \epsilon\vert x_0) = f(\tau_r^{-}\vert x_0)$ and $\lim_{\epsilon\rightarrow 0} f(\tau_r + \epsilon\vert x_0) = f(\tau_r^{+}\vert x_0)$, and recall that $\delta_t = t-n_{\tau_r}\tau_r $ with  $n_{\tau_r}=\left \lfloor \frac{t}{\tau_r} \right \rfloor$. Thus, $\delta_{\tau_r^{+}}=0$ whereas $\delta_{\tau_r^{-}} = \tau_r$. Now, since $\Lambda_t=\int_0^t\lambda_s\text{d}s$ is an integral, it is continuous and so is its Laplace transform. Thus for $t=\tau_r$, we have 
$\tilde{\Lambda}_{\tau_r^{+}}\left(1-\eta_p\right)=\tilde{\Lambda}_{\tau_r^{-}}(1-\eta_p)=\tilde\Lambda_t(1-\eta_p)=\tilde{\lambda}_e(1-\eta_p)$. 
 It follows that the discontinuity in the density given by Eq. (7) in the main text 
\begin{eqnarray}
f(t\vert x_0) = \frac{2\bar\lambda}{L}\sum_{x=1}^{L-1}\sum_{p=0}^{L}\frac{u_p(x_0,x)(1-\eta_p)\tilde\Lambda_t(1-\eta_p)}{1+(1-\eta_p)\delta_t\bar\lambda }.
\end{eqnarray}
is coming from $\delta_t$. Indeed $\delta_{\tau_r^{+}}=0$ whereas $\delta_{\tau_r^{-}} = \tau_r$ therefore 
\begin{eqnarray}
f(\tau_r^{+}\vert x_0) = \frac{2\bar\lambda}{L}\sum_{x=1}^{L-1}\sum_{p=0}^{L}u_p(x_0,x)\tilde\lambda_e(1-\eta_p)(1-\eta_p),
\end{eqnarray}
and
\begin{eqnarray}
f(\tau_r^{-}\vert x_0) = \frac{2\bar\lambda}{L}\sum_{x=1}^{L-1}\sum_{p=0}^{L}\frac{u_p(x_0,x)\tilde{\lambda}_e(1-\eta_p)(1-\eta_p)}{1+(1-\eta_p)\tau_r\bar\lambda}.
\end{eqnarray}

\noindent Finally, note that the difference between $f(\tau_r^{+}\vert x_0)$ and $f(\tau_r^{-}\vert x_0)$ vanishes for $\tau_r=0$ and becomes larger as $\tau_r$ is increased.  

A Taylor expansion further elucidates this phenomenon, demonstrating that the amplitude of the jump is directly proportional to $\tau_r$ in the case of small relaxation times. To see this we observe that  

\begin{eqnarray}
f(n\tau_r^{-}\vert x_0) &=& \frac{2\bar\lambda}{L}\sum_{x=1}^{L-1}\sum_{p=0}^{L}\frac{u_p(x_0,x)(1-\eta_p)}{(1+(1-\eta_p)\tau_r\bar\lambda)^{n+1}}.\\
&\simeq&  \frac{2\bar\lambda}{L}\sum_{x=1}^{L-1}\sum_{p=0}^{L}u_p(x_0,x)(1-\eta_p)-(n+1)\tau_r\bar\lambda^2 \frac{2}{L}\sum_{x=1}^{L-1}\sum_{p=0}^{L}u_p(x_0,x)(1-\eta_p)^2,
\end{eqnarray}

\noindent and 

\begin{eqnarray}
f(n\tau_r^{+}\vert x_0) &=& \frac{2\bar\lambda}{L}\sum_{x=1}^{L-1}\sum_{p=0}^{L}\frac{u_p(x_0,x)(1-\eta_p)}{(1+(1-\eta_p)\tau_r\bar\lambda)^{n}}\\
&\simeq&  \frac{2\bar\lambda}{L}\sum_{x=1}^{L-1}\sum_{p=0}^{L}u_p(x_0,x)(1-\eta_p)
-n\tau_r\bar\lambda^2 \frac{2}{L}\sum_{x=1}^{L-1}\sum_{p=0}^{L}u_p(x_0,x)(1-\eta_p)^2,
\end{eqnarray}
therefore 

\begin{eqnarray}
f(n\tau_r^{+}\vert x_0)-f(n\tau_r^{-}\vert x_0) &\simeq& \tau_r\bar\lambda^2 \frac{2}{L}\sum_{x=1}^{L-1}\sum_{p=0}^{L}u_p(x_0,x)(1-\eta_p)^2 \\
 &=& A(x_0,L,\bar\lambda) \tau_r
\end{eqnarray}

\noindent with $A(x_0,L,\bar\lambda)$ standing as a constant that depends on the initial location of the random walker, the size of the interval and the mean fluctuating jump rate.

%%%%%%%%%%%%%%%%%%%%%%%%%%%%%%%%%%%%%%%%%%%%%%%%%%%%%%%%%%%%%%%%%%%%%%%%%%%%%%%%%%%%%%%%%%%%%

\section{Dependence of the Mean First Exit Time on the Rate Relaxation Time}

\noindent Here, we extend the analysis in the main text regarding the dependence of the mean first exit time (MFET) on the rate relaxation time. To this end, we consider the same DSCTRW that was considered in the main text, but generalize the equilibrium jump rate distribution from exponential to Gamma. For clarity, the analytical form of the Gamma distribution is given by
\begin{equation}
  f(\lambda;\alpha, \lambda_0) = \frac{{\lambda^{\alpha - 1} e^{-\lambda/\lambda_0}}}{{\lambda_0^\alpha \Gamma(\alpha)}}  
\end{equation}
\noindent Here, $\lambda$ represents the equilibrium jump rate, $\alpha$ is the shape parameter, and $\lambda_0$ is the rate parameter. Note that at small jump rates the Gamma distribution obeys the following power-law scaling $\sim \lambda^{\alpha-1}$. In what follows, we explore three distinct Gamma distributions: $\Gamma(\alpha=1,\lambda_0=1)$, $\Gamma(1/2,2)$, and $\Gamma(2,1/2)$. Note that in the three examples we consider the mean is fixed and given by: $\langle\lambda\rangle=\lambda_0\alpha=1$. However, since $\alpha$ is different---taking values both above and below unity---the behaviour of these distribution at small jump rates is also very different. Next, we show that this has a drastic impact on the asymptotic dependence of the MFET on the rate relaxation time.

\begin{figure}[!t]
    \centering
    \includegraphics[width=12cm]{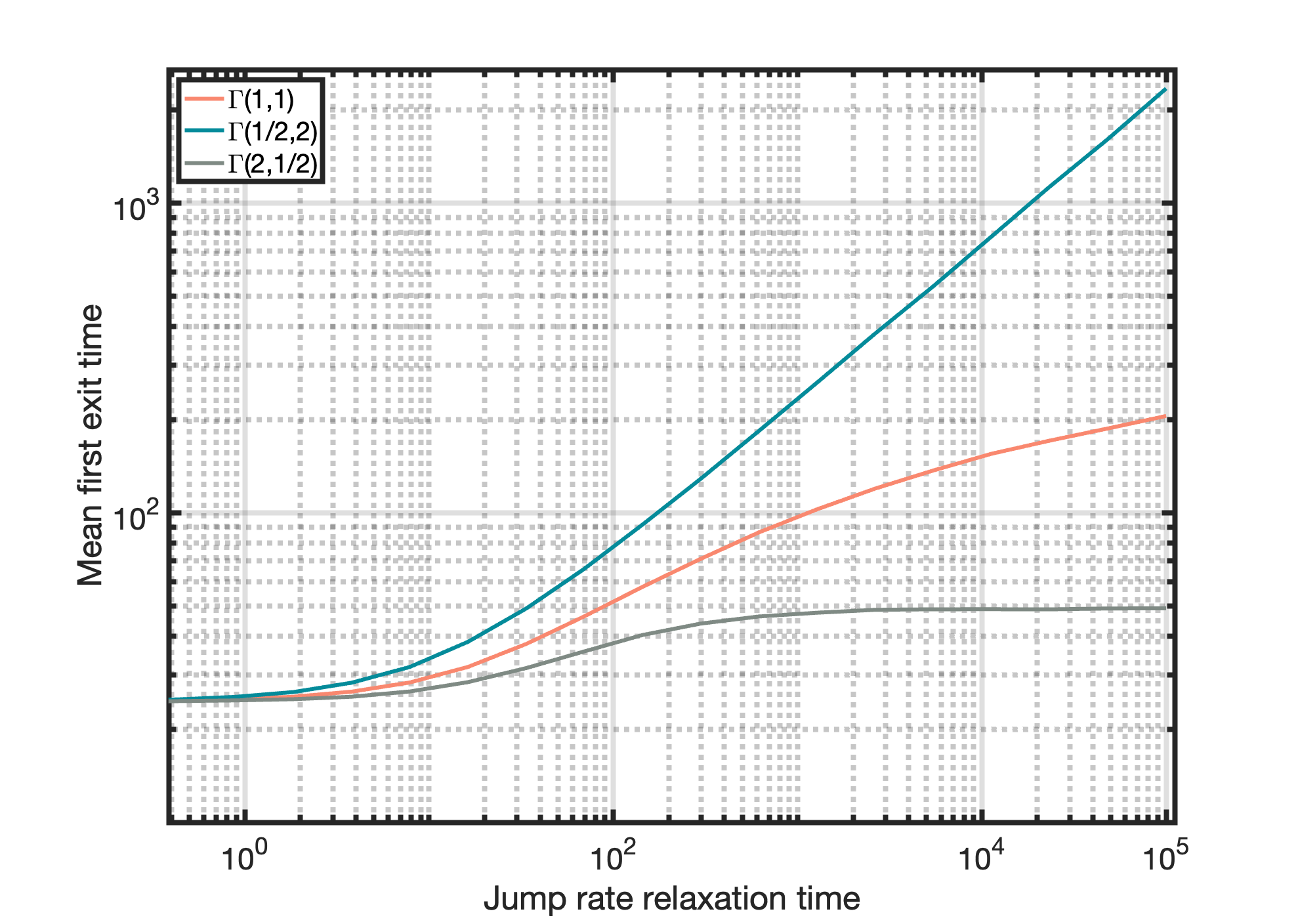}
    \caption{Mean first exit time (MFET) from the interval $\left[0,10\right]$, starting  at $x_0=5$, for the DSCTRW illustrated in Fig.~1 of the main text, albeit with jump rate whose steady-state distribution is Gamma (thus generalizing the exponential). Three distinct cases are depicted: $\Gamma(\alpha=1,1)$ (orange), $\Gamma(\alpha=1/2,2)$ (cyan)), and $\Gamma(\alpha=2,1/2)$ (gray). The MFET is plotted against the jump rate relaxation time ($\tau_r$) on a log-log scale. For fast jump rate relaxation (small $\tau_r$), we recover the MFET of a simple CTRW with a fixed jump rate corresponding to the mean jump rate, which was here set to unity. In the other extreme, i.e., for slow relaxation times (large $\tau_r$), the behaviour of the MFET is governed by the behaviour of the jump rate distribution at small jump rates. For the Gamma distribution considered here the jump rate distribution at small jump rates obeys a power-law scaling $\sim \lambda^{\alpha-1}$. For $\alpha=1$ (orange curve) this leads to a logarithmic divergence of the mean first exit time at large $\tau_r$, while for $0<\alpha<1$ (cyan curve) we find a $\sim \tau_r^{\alpha}$ power-law divergence . In contrast, for $\alpha>1$ (gray curve), we the mean first exit time saturates to a constant at large relaxation times.}
    \label{fig:mfpt_dsctrw}
\end{figure}

\noindent In the limit of short relaxation times, the MFET (for the example considered in the main text) can be approximated as follows
\begin{equation}
    \langle T_{DSCTRW}\rangle = \frac{x_0(L-x_0)}{\langle\lambda\rangle} = \frac{x_0(L-x_0)}{\lambda_0\alpha} = \langle T_{CTRW}\rangle
\end{equation}
\noindent We thus see that the MFET in this limit is equal for three examples and given by $x_0(L-X_0)=25$. To determine the behaviour at the large  relaxation time limit, one must consider the average of $1/\lambda$. Notably, when $\alpha = 1$, the Gamma distribution simplifies to the exponential distribution with rate parameter $\lambda_0$, corresponding to the case $\Gamma(1,\lambda_0)$. In this scenario, whose MFET is represented by the orange curve in the Fig. S4, the MFET diverges as $\ln(\tau_r)$,  This divergence can be understood by averaging the MFET of a regular CTRW with jump rate $\lambda$ and using the distribution of $\lambda$ with a cutoff at $1/\tau_r$ to prevent divergence of the integral
\begin{equation}
    \langle T_{DSCTRW}\rangle = \int_{\tau_r^{-1}}^{+\infty}\frac{x_0(L-x_0)}{\lambda}\frac{e^{-\lambda/\lambda_0}}{\lambda_0}\text{d}\lambda= T_{CTRW}\int_{\tau_r^{-1}}^{+\infty}\frac{e^{-\lambda/\lambda_0}}{\lambda}\text{d}\lambda \propto T_{CTRW}\ln(\tau_r).
\end{equation}
\noindent The $\Gamma(1/2,2)$ case, whose MFET is represented by the cyan curve in the Fig. S4, exhibits a divergence of the jump rate distribution at zero. This implies that the mean exit time becomes infinitely large as $\tau_r$ increases. Using the integral above, we find that the MFET diverges according to 
\begin{equation}
    \langle T_{DSCTRW}\rangle = \int_{\tau_r^{-1}}^{+\infty}\frac{x_0(L-x_0)}{\lambda}\frac{e^{-\lambda/2\lambda_0}}{4\lambda_0^2\sqrt{\pi\lambda}}\text{d}\lambda= \frac{T_{CTRW}}{4\lambda_0\sqrt{\pi}}\int_{\tau_r^{-1}}^{+\infty}\frac{e^{-\lambda/2\lambda_0}}{\lambda^{3/2}}\text{d}\lambda \propto T_{CTRW}\sqrt{\tau_r},
\end{equation}
\noindent On the other hand, the $\Gamma(2,1/2)$ distribution vanishes at zero sufficiently rapidly. As the integral converges when $\tau_r$ goes to infinity, the MFET will saturate to a finite value, as illustrated by the gray curve in Fig. S4.

\bibliographystyle{plain}
\bibliography{biblio.bib}